# Biomimetic surface structuring using cylindrical vector femtosecond laser beams

Evangelos Skoulas[1], Alexandra Manousaki[1], Costas Fotakis[1] and Emmanuel Stratakis[1,2]*

[1]Institute of Electronic Structure and Laser (IESL), Foundation for Research and Technology (FORTH), N. Plastira 100, Vassilika Vouton, 70013, Heraklion, Crete, Greece
[2] Materials Science and Technology Department, University of Crete, 71003 Heraklion, Greece
*Corresponding author: stratak@iesl.forth.gr

**Abstract**

We report on a new, single-step and scalable method to fabricate highly ordered, multi-directional and complex surface structures that mimic the unique morphological features of certain species found in nature. Biomimetic surface structuring was realized by exploiting the unique and versatile angular profile and the electric field symmetry of cylindrical vector (CV) femtosecond (fs) laser beams. It is shown that, highly controllable, periodic structures exhibiting sizes at nano-, micro- and dual-scale micro/nano scales can be directly written on Ni upon line and large area scanning with radial and azimuthal polarization beams. Depending on the irradiation conditions, new complex multi-directional nanostructures, inspired by the *Shark's* skin morphology, as well as superhydrophobic dual-scale structures mimicking the *Lotus'* leaf water repellent properties can be attained. It is concluded that the versatility and features variations of structures formed upon scanning with CV beams is by far superior to those obtained via laser processing with linearly polarized beams. More important, by exploiting the capabilities offered by fs CV optical fields, the present technique can be further extended to fabricate even more complex and unconventional structures. We believe that our approach provides a new concept in laser processing of materials, which can be further exploited for expanding the breadth and novelty of potential applications.

# 1. Introduction

Nature has always provided a plethora of functional surfaces exhibiting unique, complex hierarchical morphologies with dimensions of features ranging from the macroscale to the nanoscale. Such morphologies are always behind the superior properties exhibited by the natural surfaces, including extreme wetting, floatation, adhesion, friction and mechanical strength [1]. In this context, the design and the fabrication of biomimetic structures is of significant importance and provides a virtually endless potential for the development of novel artificial materials and systems.

Despite the increasing scientific interest, the complex structure of most of the natural surfaces has been proven to be extremely difficult to mimic. Several fabrication techniques have been developed, based on top-down and bottom-up processing schemes [1]. Among those, direct laser structuring is a material independent and versatile technique that presents some key benefits for precise surface modification, over competitive techniques [2]. In particular, laser processing with fs pulses offers advantages in minimizing thermal effects and collateral damage, allowing localized modifications with a large degree of control over the shape, size and the range of features that can be produced. Indeed, fs laser induced surface structuring has been demonstrated to produce numerous biomimetic structures [2–5] for a range of applications, including microfluidics [3,5,6], tribology [7–9], tissue engineering [2,10] and advanced optics [11].

A prominent aspect of the fs laser material interaction is that the spatial features of the surface structures attained are strongly correlated with the laser beam polarization. This is for example the case of laser-induced periodic surface structures (LIPSS) and quasi-periodic microgrooves, which are preferentially oriented perpendicular and parallel to the laser polarization respectively [12–17]. However, to date, laser fabrication of biomimetic structures has been demonstrated using laser beams with a Gaussian intensity spatial profile and spatially homogeneous linear polarization [18]. In this context and based on the sensitivity of laser induced structures on laser polarization, it is possible to further advance the complexity of the fabricated structures via utilizing laser beams with a spatially inhomogeneous state of polarization [19–22]. CV beams, exhibiting tangential polarization states, are prominent examples [23–28].

In this paper, we report on the direct fs laser biomimetic surface structuring via the use of CV beams generated by an s-waveplate [29], which transforms a linearly polarized Gaussian beam to a CV beam with tangential polarization states. It is shown that dynamic surface processing with radial and azimuthal polarization beams, gives rise to large areas of complex biomimetic structures. In particular, the formation of bioinspired multi-directional periodic structures that mimic the sharks' skin morphology, as well as that of well-ordered superhydrophibic hierarchical micro/nano structures on Ni surfaces, is demonstrated. Although the fabrication of these particular morphologies was demonstrated, the versatility and features' variations of attainable structures could be practically endless. We believe that our approach brings about a new thinking in laser processing of materials and can be further extended to provide even more complex biomimetic structures for numerous potential applications.

## 2. Experimental section

The experimental apparatus used to fabricate biomimetic surface structures using CV beams is presented in Fig.1. Commercially available polished Ni films of 99.9% purity and average thickness of 100μm where used as samples. The Yb:KGW laser source produced linearly polarized pulses of 170fs, 1KHz repetition rate and 1026nm central wavelength, while the Gaussian spot diameter, measured by a CCD camera at $1/e^2$, was 32μm. CV beams of radial and azimuthal polarisation, exhibiting a donut-shaped profile, were generated by means of an s-waveplate. The characteristics of the produced CV beams are presented in Fig. 5S. Following the s-waveplate, the CV beam was focused on the sample via an achromatic convex lens of 60mm focal length. Samples were fixed onto a 3-axis motorized stage and positioned perpendicular to the incident beam. All irradiations were performed in ambient environment. Due to the different spatial profile of the Gaussian with respect to an CV beam, fluence, $\varphi$ calculations were made separately in each case as shown at [24,25]. At the same time, the incident number of pulses, *NP*, was controlled by an electromagnetic beam shutter.

For the dynamic processing experiments, line or area scans were produced at variable velocity values, *v*, ranging from 0.3mm/s to 2.0mm/s. In this case, the effective number of laser pulses ($N_{eff}$) per unit length or area should be determined respectively. $N_{eff}$ has been commonly used for Gaussian beams and corresponds to the number of laser pulses falling, upon one-dimensional scanning, onto a length interval equal to the Gaussian beam diameter $2w_0$. In the case of CV beams, $N_{eff}$ should depend on the corresponding donut shaped beam area. For line scanning at constant velocity *v* and at repetition rate *f* and assuming donut diameters *R* (outer), *r* (inner), the effective pulse number $Neff_{line}$ can be defined as:

$$Neff_{line} = 2(R-r) \cdot \frac{f}{v} \quad (1)$$

While for large area scanning at constant velocity *v*, at repetition rate *f* and individual line separation *δ*, $Neff_{area}$ can be defined as:

$$Neff_{area} = \pi(R^2 - r^2) \cdot \frac{f}{v \times \delta} \quad (2)$$

While, the spot overlap area is defined as:

$$Overlap_{area} = 2R^2 \cos^{-1}\left(\frac{d}{2(R)}\right) - \frac{d}{2}\sqrt{4(R)^2 - d^2} \quad (3)$$

$$d = \frac{v}{f} \quad (4)$$

Where *d* is the distance between two consecutive circular spot centers.

The morphology of the laser-induced structures has been characterized by scanning electron microscopy (JEOL JSM-7500F). While, the structures' characteristics was determined by two dimensional fast Fourier transform (2D-FFT) analysis of the respective SEM images using the Gwyddion software. Details for the periodicity calculation can be found in the supporting information Fig1S. For higher accuracy and error estimation, SEM images of three spots, produced with identical conditions were statistically analyzed.

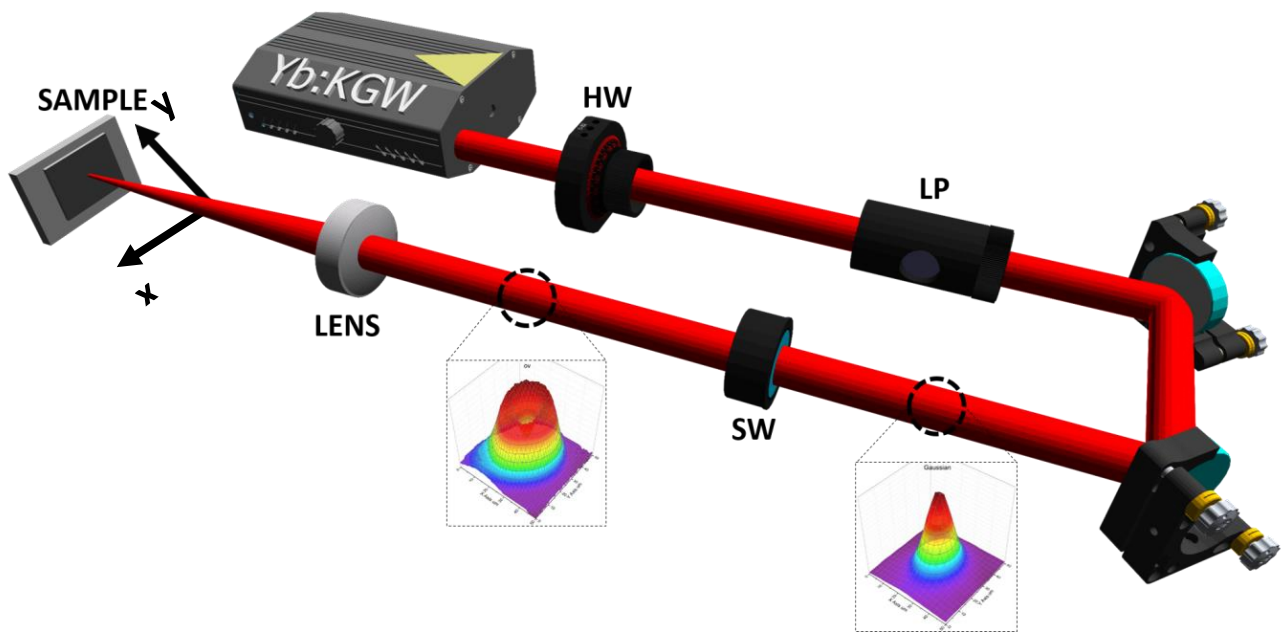

**Figure 1.** Schematic of the experimental setup developed for the laser induced fabrication of biomimetic structures. The incident laser fluence was varied by means of a λ/2 waveplate (HW). The Gaussian profile emitted by the laser source is transformed to a CV beam using a rotating s-waveplate (SW). Depending on the SW rotation angle, the CV beam polarization can be changed from radial to azimuthal respectively.

The wetting properties of the fabricated surfaces were measured by the sessile droplet method, performed using the DataPhysics OCA 20 system. In particular, distilled water drops of 4μl were deposited on each surface tested and the average value of the water contact angle (CA), obtained from five measurements as well as the standard deviation, was calculated.

## 3. Results

### 3.1 Single and multiple shot irradiation experiments

In a first step, the characteristics of laser-induced structures formed upon variation of the *NP* (1 - 1000 at a constant fluence of 0.24J/cm$^2$), as well as incident fluence (0.17J/cm$^2$ - 0.74J/cm$^2$, at a constant *NP*=100) of the CV beams, was investigated. SEM imaging of the respective spots indicated that for *NP*<5 no structures were formed in the whole range of fluences used. While, from 2≤*NP*≤5, surface roughness was significantly increased and resembles a nanostructured grating with a tendency to orientate parallel to the incident polarization (Fig. S2). For higher *NP*, the resulting surface comprises a central microstructure formed in the inner region of the CV beam, exhibiting almost null intensity, while the donut area was always decorated with a characteristic texture of LIPSS, always aligned perpendicular to the laser polarization. Accordingly, LIPSS produced with azimuthally polarized light showed radial spatial distribution, while LIPSS obtained with radial polarization exhibited a concentric ring spatial distribution. Fig. 2 presents typical examples of SEM images of such structures for Gausian linear, CV radial and azimuthal polarization respectively at specific irradiation conditions. The results from the parametric analysis, described above, showed that, regardless the polarization condition, the LIPSS periodicity decreases with *NP*s, while it is weakly influenced by the incident fluence (Fig. 3). At the same time, the crater depth and thus the height of the microstructure formed at the spot center of the CV beam can be changed, upon increasing NP and/or laser fluence, in the range from hundreds of nanometers to a few tens of microns.

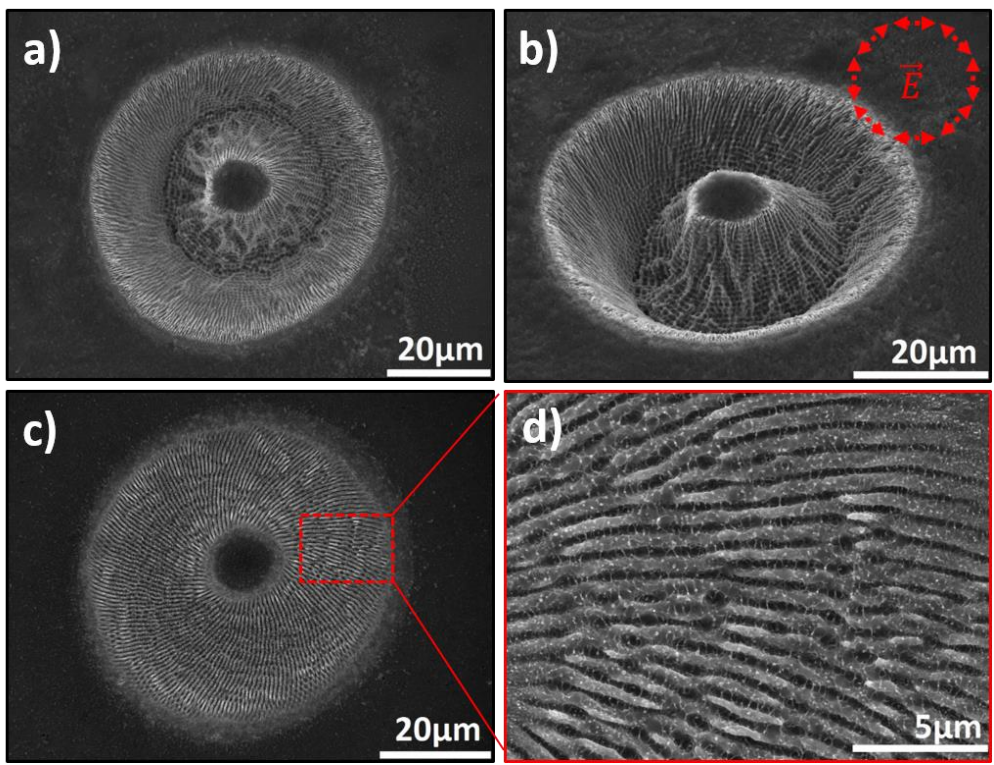
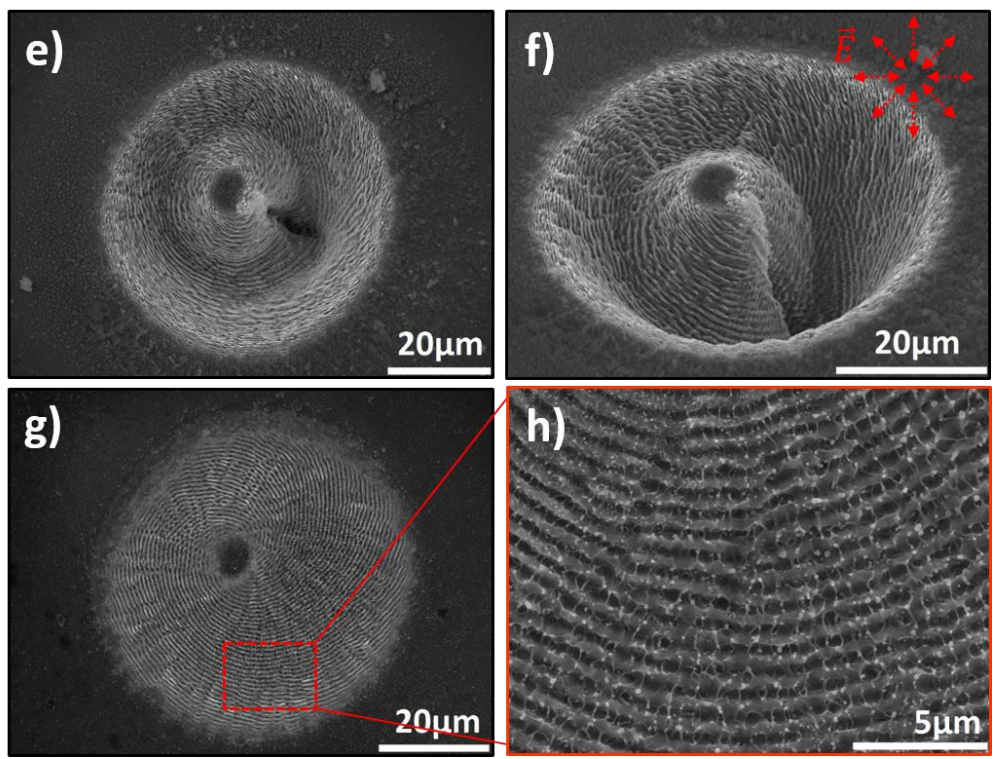

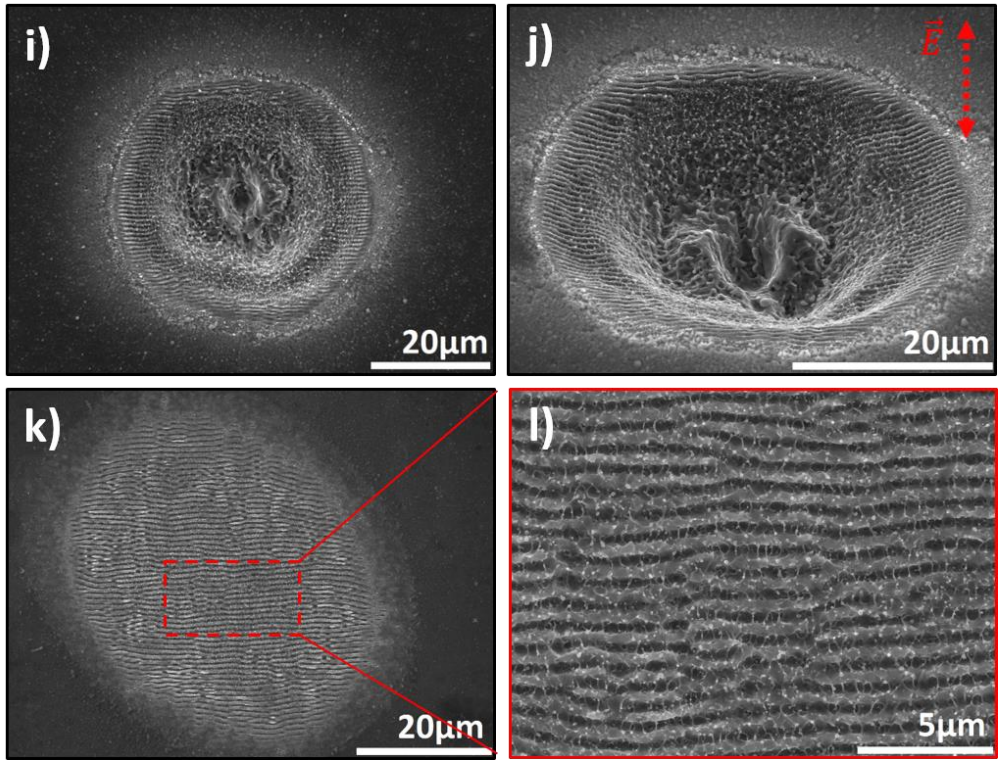

**Figure 2.** SEM images of fs laser-induced structures formed on Ni surfaces upon irradiation with azimuthal (a)-(d), radial (e)-(h) and linear (i)-(l) polarization beams respectively, using φ=0.24J/cm$^2$ and NP=1000 (a,b,e,f,I,j) or NP=100 (c,d,g,h,k,l). All pictures show top-views, except b), f), j) that present 45-degrees views. Images d), h) and l) are higher magnifications of the red-dashed-square areas.

As shown in Fig. 3, the LIPSS period progressively decreases for 5 < NP < 600, with a trend to saturate at higher NPs, an effect which is valid for both Gaussian (linear polarization in Fig. 3) and CV beams. On the contrary, LIPSS period is weakly dependent on the incident fluence, despite the spatial profile and polarization of the beam. This behavior has been recently addressed by our group [30], via a combined theoretical and experimental study, showed that it can be attributed to a synergy of electrodynamic and hydrodynamical effects [13].

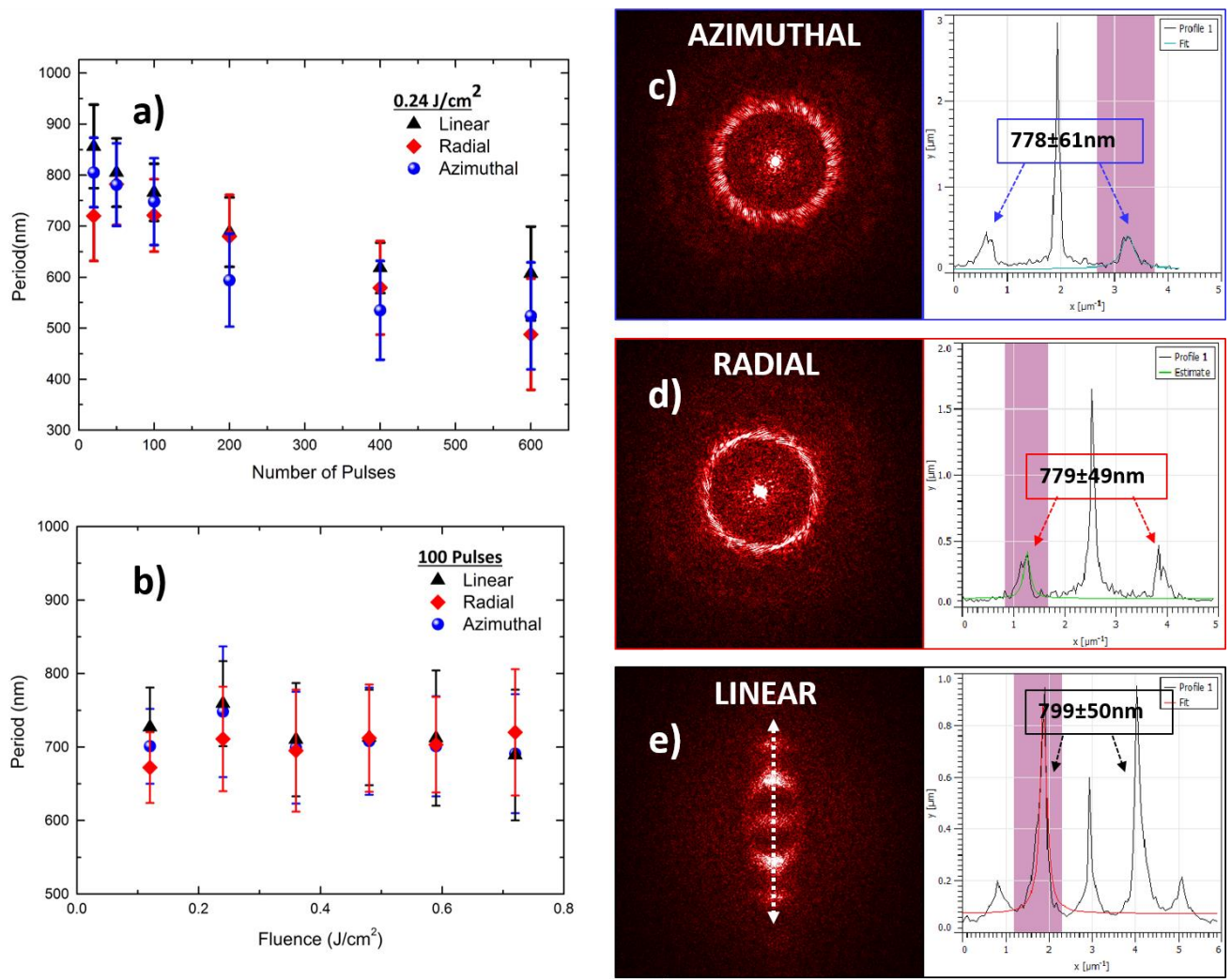

**Figure 3.** LIPSS dependence on (**a**) the fluence and (**b**) the number of pulses for the azimuthal, radial and linear beams respectively. On the right side the 2D-FFT analysis corresponding to the SEM images c), g) and k) of Fig. 2 is demonstrated. Specifically, the 2D-FFT image spectra are presented for each polarization, together with the respective line profiles obtained via image cross-sections. The corresponding calculated LIPSS periodicity is shown in the inset.

Fig. 3(c) – (e) presents also the 2D-FFT analysis used to determine the ripples periodicity, performed on typical SEM images of spots created by linear Gaussian, radial and azimuthal CV beams respectively. It can be observed that the 2D-FFT image characteristics reveal the polarization type, as it exhibits a preferential directionality for linear, a radial distribution for azimuthal and a vortex-like pattern for radial polarization respectively.

### 3.2 Line scans using CV beams

Following spot analysis, line processing experiments were performed in scanning mode, using different scan velocities (thus $N_{eff,line}$, calculated by Eq. 3) and spot overlap at a constant $\varphi$ value. Fig. 4A exemplifies the characteristic surface morphologies attained, in top-view SEM micrographs of scanned lines, obtained at $v$=0.5mm/s ($Neff_{line}$=62) and $\varphi$=0.24J/cm$^2$, for linear Gaussian (4A-a, b), radial (4A-c, d) and azimuthal (4A-e, f) CV beams respectively. Contrary to the, mostly applied to date, linear polarization case, dynamic surface processing with CV beams produces multi-directional rhombus-like structures exhibiting a radial or azimuthal directionality respectively. Such rhombus-like structures significantly mimic the placoid structures found in the skin of shark, although their characteristic size differs substantially from those comprising the shark skin [8,31,32].

To shed light on the creation of such novel structures, the line scan areas were carefully examined by SEM imaging. As presented in Fig. 4A-g, it was revealed that the central part of the scan lines is patterned with LIPSS exhibiting a radial or azimuthal orientation (see Fig. 2), depending on the CV beam used for scanning. While, the linescan peripheral areas are textured with the shark skin-like rhomboid structures. As presented in Fig. 4B, such structures become as the natural outcome of the overlap between successive CV beam spots, specifically as the beam advances on the surface during scan, the pulse overlapping effect leads to crossed vector peripheral areas, giving rise to the rhombic-shaped structures.

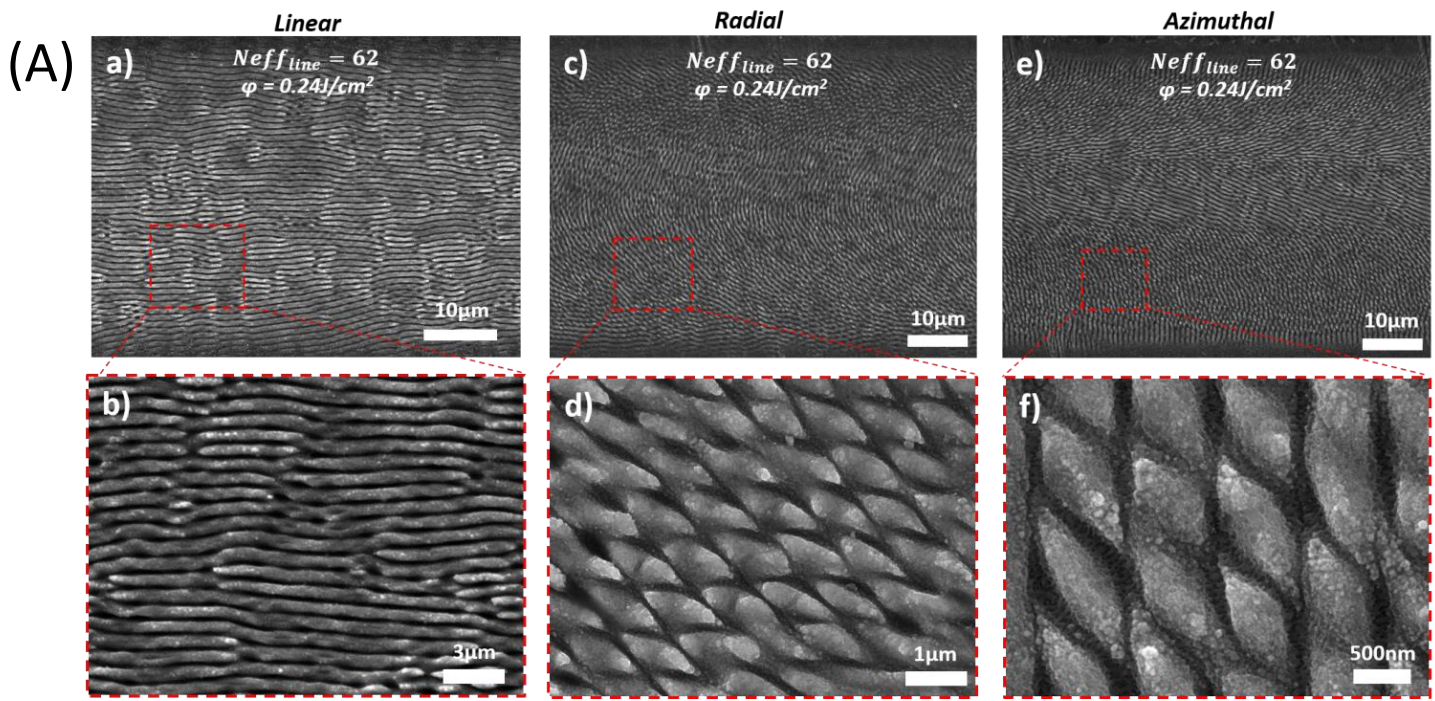

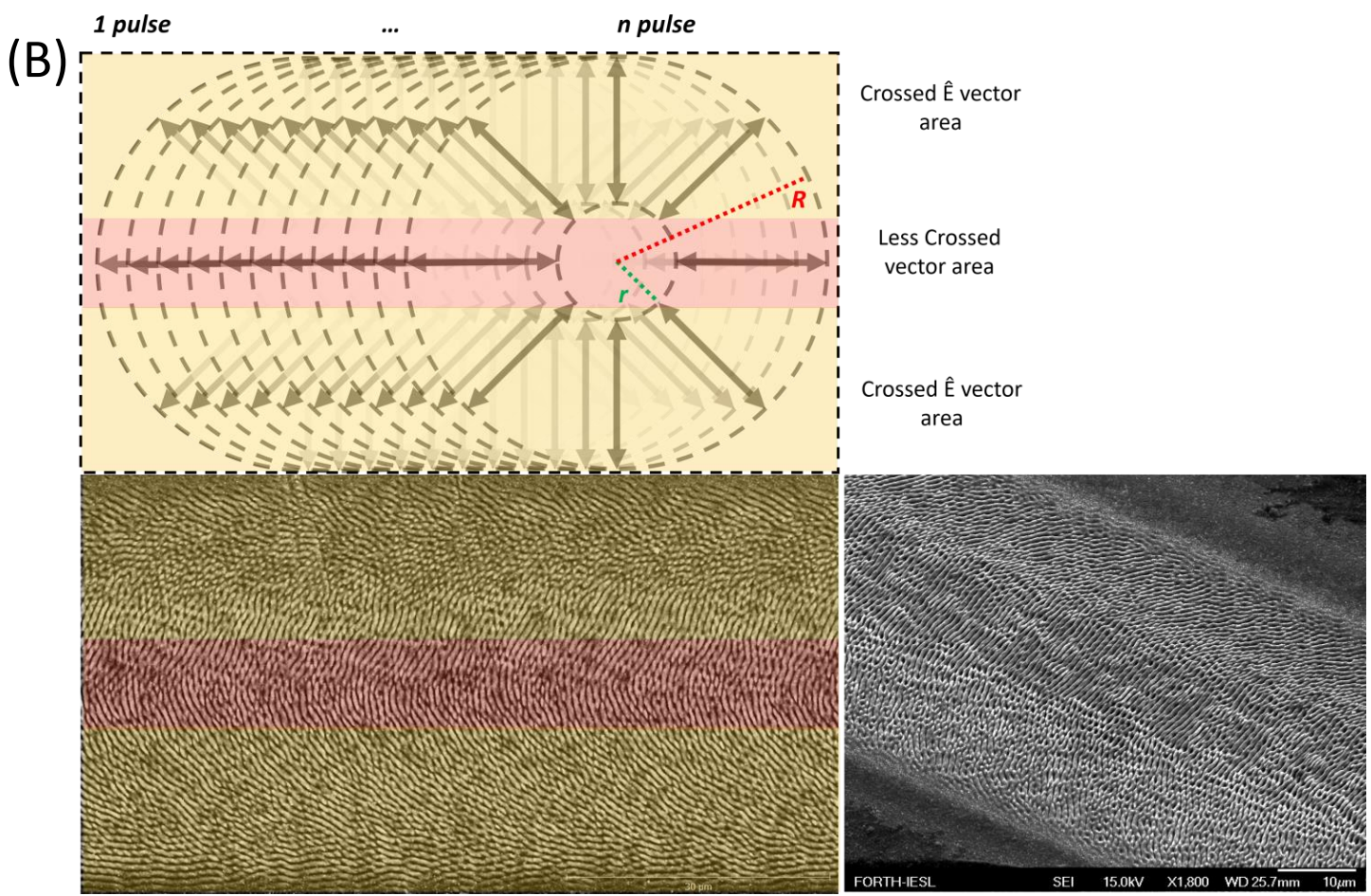

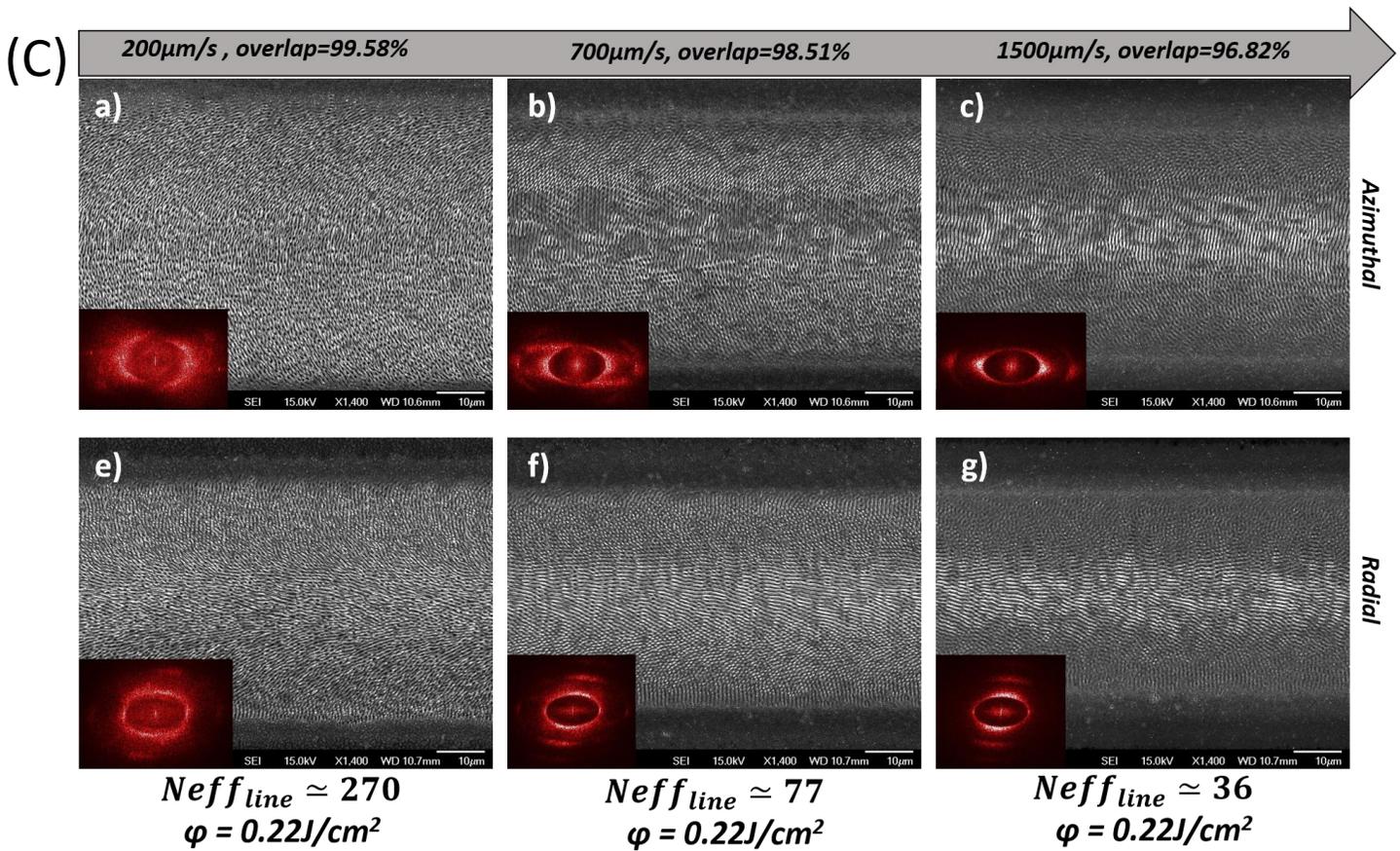

**Figure 4:** A) Top-view SEM images depicting line scans produced by linearly (a, b), radially (c, d), and azimuthally polarized (e, f) beams, respectively, at $v$=0.5mm/s ($Neff_{line}$=62), and $\varphi$=0.24J/cm². The images (b), (d), (f) are higher magnifications of the red-dashed-square areas and revel the biomimetic shark skin-like morphology of the processed areas; B) Schematic of the beam overlap process during a linescan with a CV beam; a typical linescan SEM image is shown for comparison; C) SEM images of linescans produced by azimuthally (a, d) and radially (e, h) polarized CV beams of a constant fluence, $\varphi$=0.24J/cm², at different scanning speeds. The corresponding 2D-FFT images are shown as insets.

According to the above, the scanning speed, that determines the degree of overlap, should play significant role on the patterns' morphology attained. This is evident in Fig. 4C, depicting SEM images of line scans, fabricated at variable scan velocities. The insets represents the corresponding 2D-FFT images for each line scan, showing that, as the spot overlap is increased, additional spatial frequencies are generated. It is clear that the scan speed significantly affects the complexity and thus the variety of the structures attained.

### 3.3 Fabrication of large areas using CV beams

The complexity of the structures attained can be further enriched upon areal scans with CV beams. Fig. 5a (a)-(h) presents SEM images of 16mm$^2$ areas, fabricated at $\varphi$=0.37J/cm$^2$, $v$=2mm/s and $Neff_{area}$=31, using linear Gaussian, radial and azimuthal beams respectively. It can be observed that the structures attained are multi-directional with no preferential shape and orientation. Such surface morphology occurs due to the overlap between adjacent spots during the scanning process. Accordingly, the scanning speed together with the overlap between adjacent spots, are the most important parameters affecting the morphology of the surfaces attained. The effect of the scanning speed is presented in Fig. 5b showing large areas fabricated using azimuthally polarized CV beam, at different scanning velocities. Morphological and 2D-FFT imaging analyses reveal that when the scanning speed is high (low $Neff_{area}$), the structures showed higher periodicity and a more linearly arranged orientation, closer to that observed upon processing with linearly polarized beams. On the contrary the structures fabricated with high $Neff_{area}$ (i.e. at low speeds), are characterized by many spatial frequencies; such morphologies could not be attained upon processing with linearly polarized fs beams.

The physical properties of the unique surface morphologies attained are yet to be examined and further work, in this direction, is under progress. Wettability characterization of such surfaces showed only a small increase in hydrophobicity, compared with the non-irradiated, one. At the same time, some of the surfaces exhibited a remarkable structural coloration, an example of which is presented in Fig. 5c, the exact optical properties of such unique structures are, however, under investigation.

The above findings suggest that surface processing with ultrashort CV beams could be a novel approach to increase the complexity of the structures attained and thus further advance the capabilities of ultrashort pulsed laser processing.

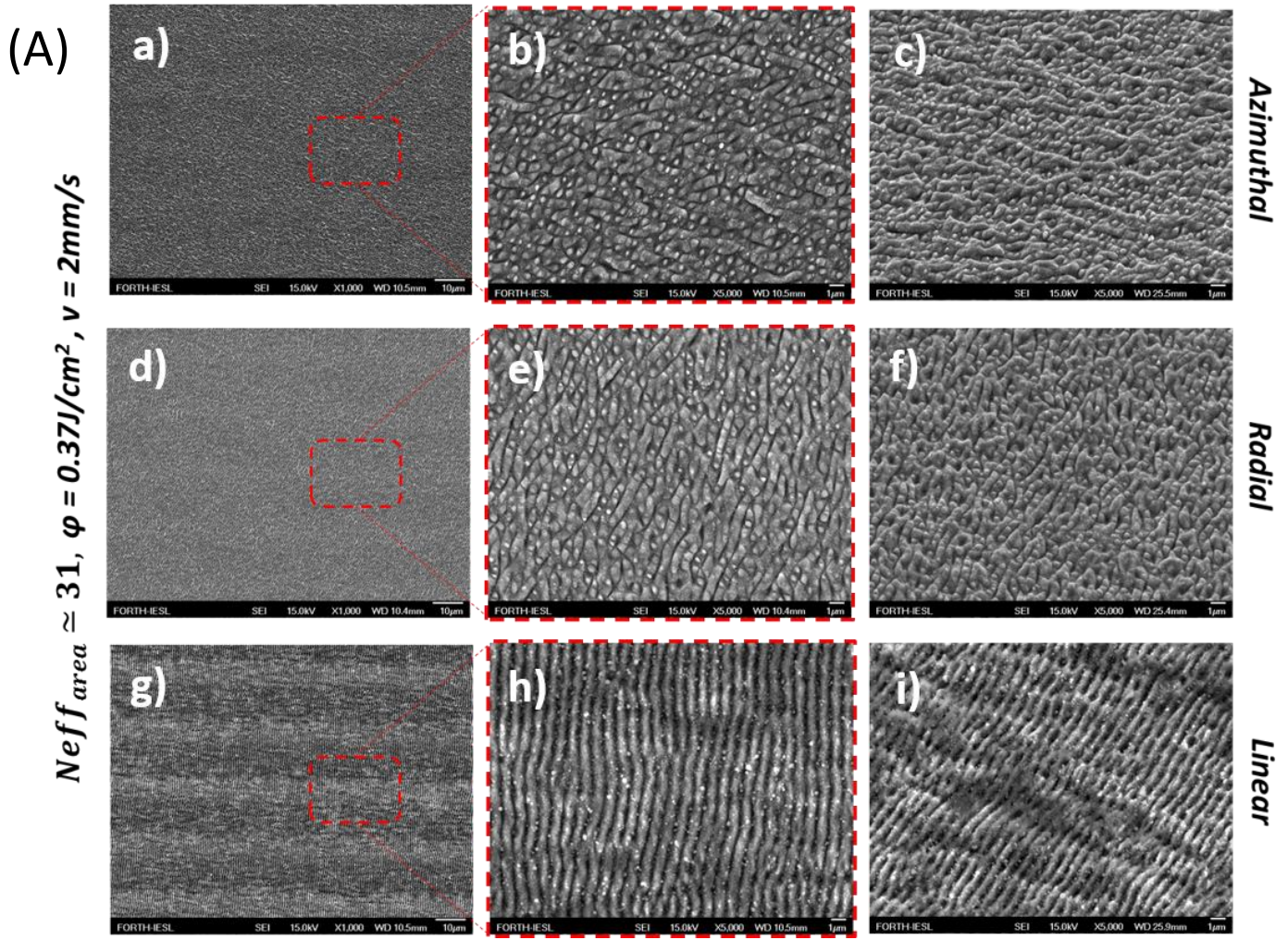

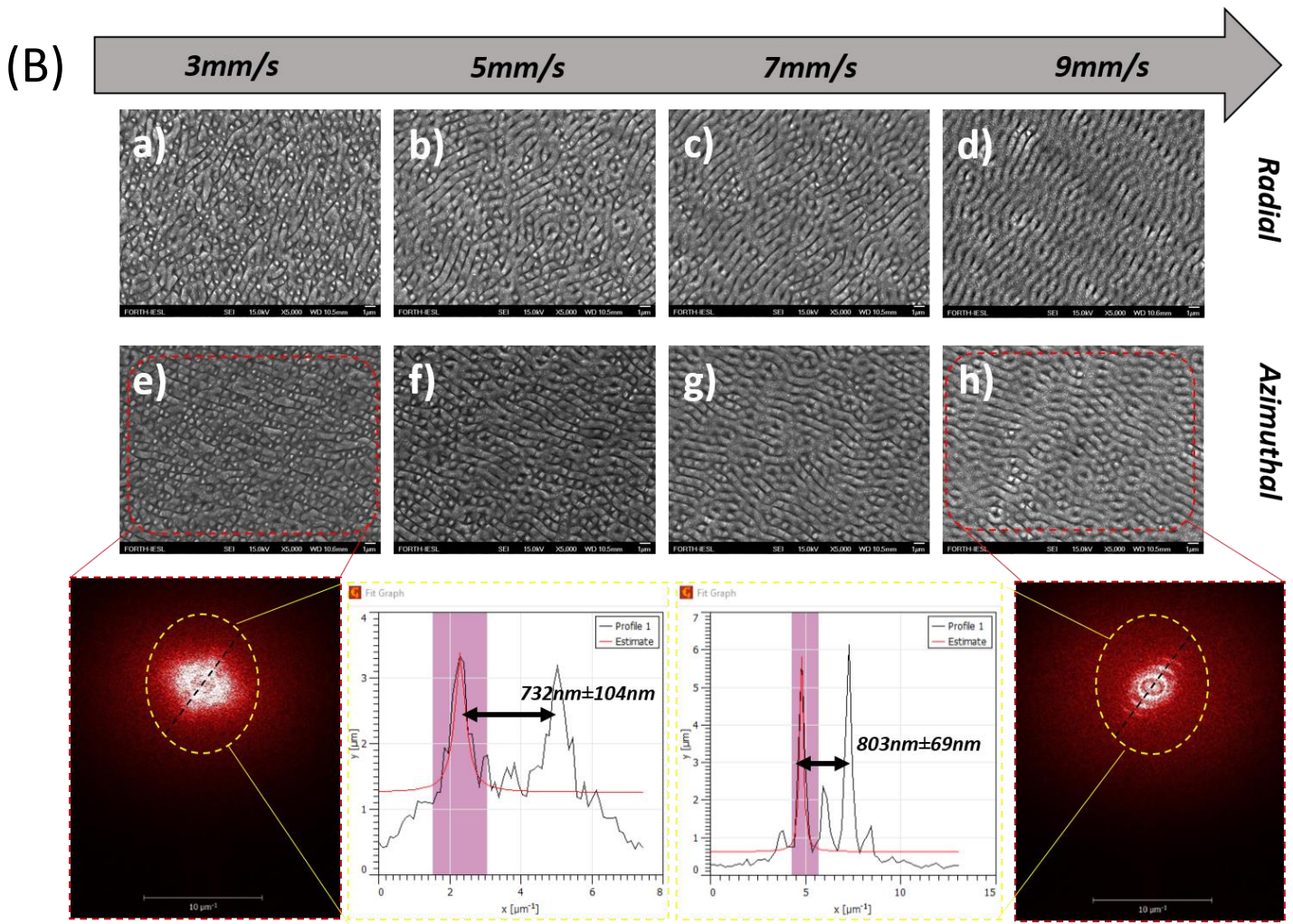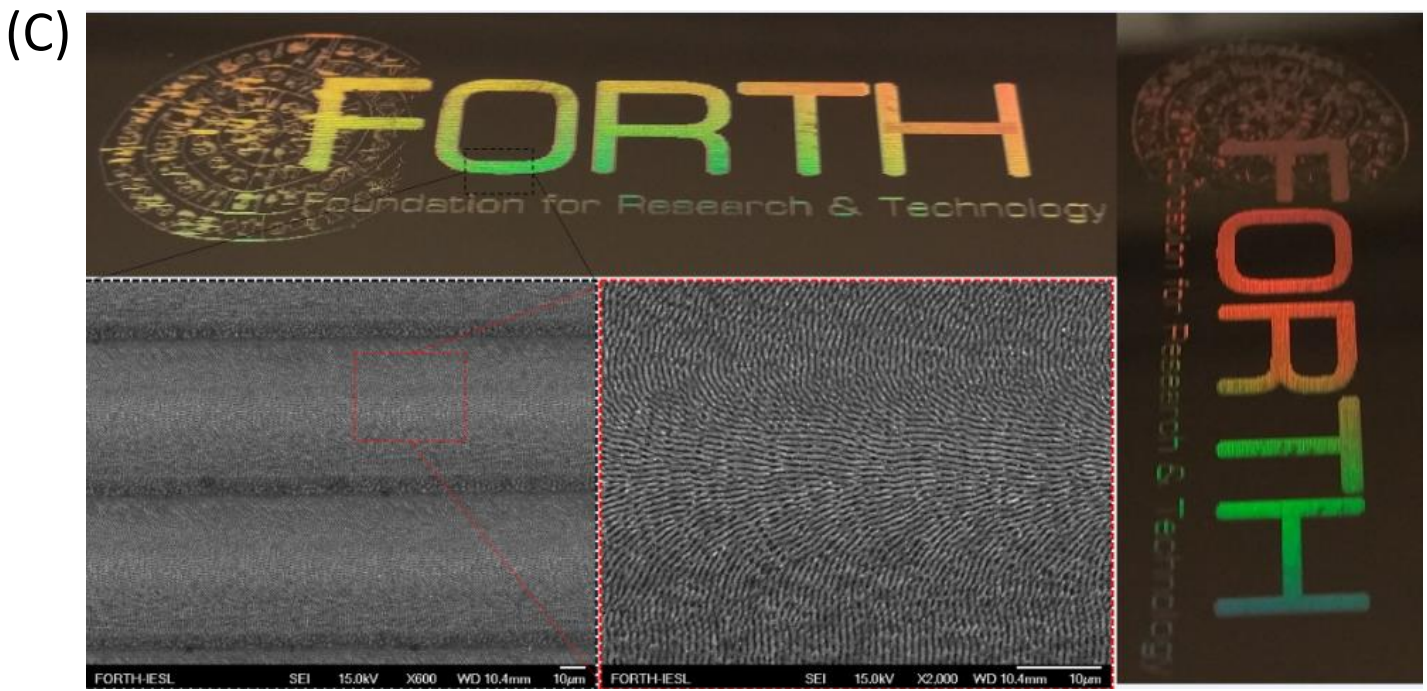

**Figure 5**: A) SEM images from 16 mm$^2$ areas fabricated using azimuthal (a-c), radial (d-f) and linear (g-i) polarization fabricated at v=2 mm/s ($Neff_{line}$=31), and $\varphi$=0.37J/cm$^2$. All pictures show top-views, except c), f) and i) that present 45-degrees views. Images b), d) and h) are higher magnifications of the red-dashed-square areas. B) SEM images of area scans produced by radially (a-d) and azimuthally (e-h) polarized CV beams of a constant fluence, $\varphi$=0.24J/cm$^2$, at different scanning speeds. The corresponding 2D-FFT images for the lower and the higher scanning speed are shown, together with the respective 2D-FFT line profiles. The characteristic period of the ripple-like structures are also presented; C) Structural coloration exhibited by a large area pattern, depicting the FORTH logo, fabricated with a azimuthally polarized CV beam of $\varphi$=0.36J/cm$^2$ at v=1 mm/s ($Neff_{line}$=30).

**3.4 Fabrication of large areas of hierarchical structures.** A multi-scale structuring approach is especially important in mimicking natural surfaces comprising structures with morphology at multiple length scales. In this respect, the fabrication of precisely controlled surfaces consisting of structures with more than one spatial frequency, is desirable. As mentioned above and presented in Figure 1, the crater profiles corresponding to multiple pulse irradiation with CV beams, exhibit a primary microstructure at the center decorated with secondary submicron ripples. This morphology could provide a template motif for the fabrication of dual-scale, high- and low- spatial frequency structures. In a first step, a sequence of CV spots (Fig. 6) was used to fabricate a periodic array of microcones (the height of which, can be varied upon changing the incident φ), complemented by LIPSS (the periodicity of which can be tuned, in the range 1000-400nm, upon changing the incident *NP*). At the same time, the distance among microstructures can be varied upon changing the overlap between adjacent spots. Using such a spot by spot process, areas of several millimeters square were fabricated. A typical example is presented in Fig. 6, showing a top-view SEM image of an array of microcones, decorated with periodic ripples of average periodicity of about 500 nm, produced by a series of radially polarized spots repeated over the whole area with overlap of 15.9% between two concecutive spots.

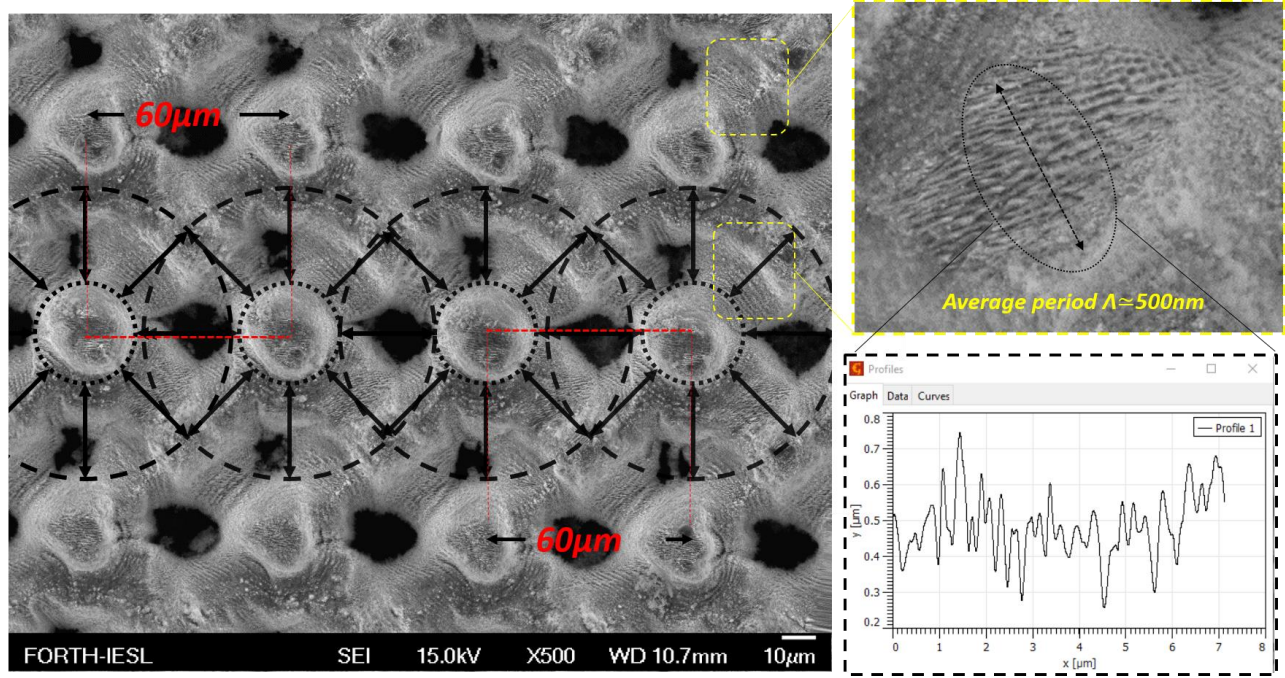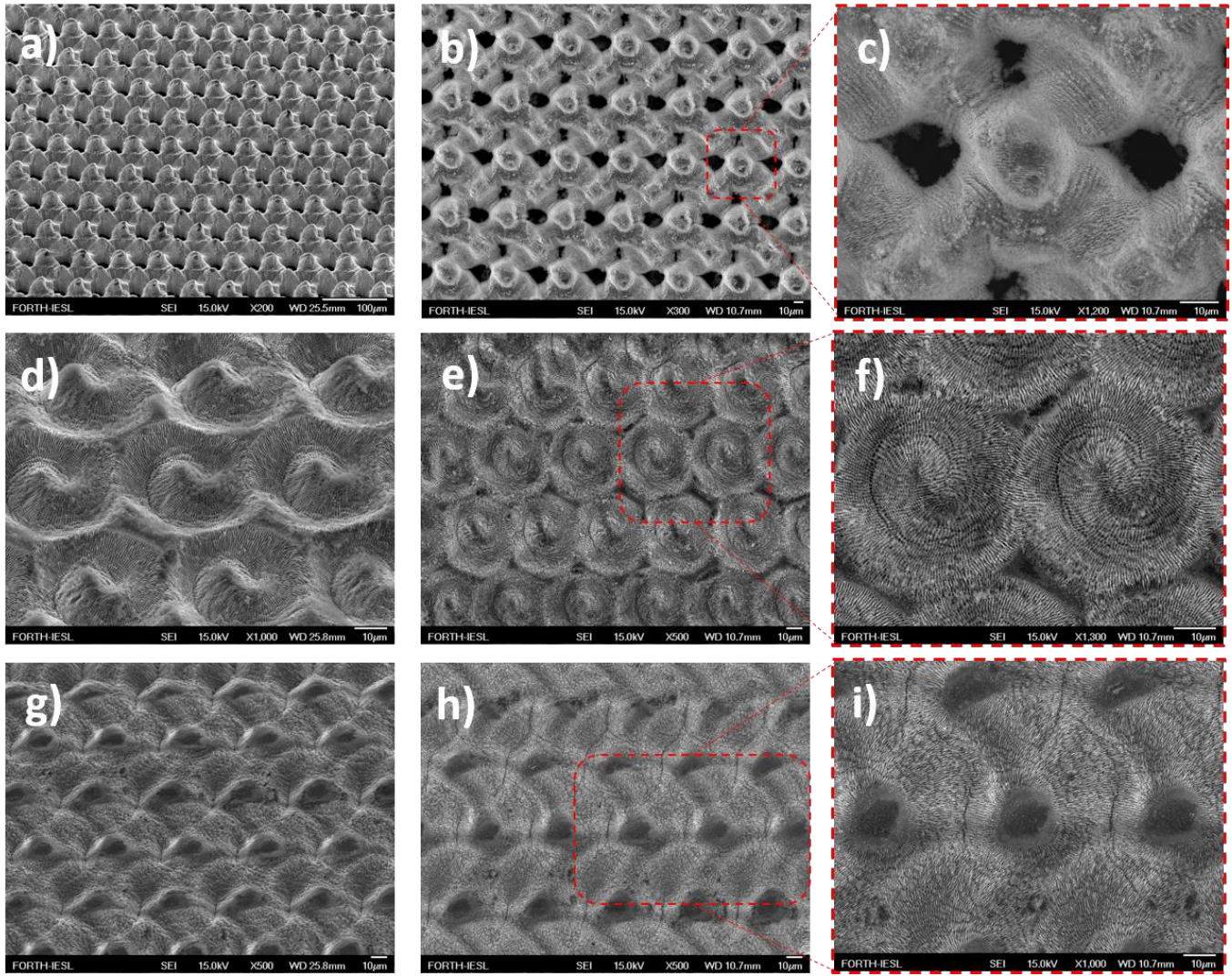

**Figure 6.** A) Top-view SEM image of a surface fabricated via spot-by-spot processing 500 radially polarized pulses of 9.45J/cm$^2$ per spot; the dashed lines present the corresponding beam overlap upon spot-by-spot processing; the submicron sized rippled-like secondary structure decorating the primary microstructures is shown on the right, together with a respective line profile used to calculate the periodicity; B) SEM images of three types of dual-scale surfaces fabricated via spot-by-spot processing with CV beams; (a,b) show low- and high- magnification 45$^0$ views of HR surfaces fabricated with 500 radially polarized pulses of 9.45J/cm$^2$ per spot; (c) shows the top view of a single microstructure of the HR surface; (d,e) show low- and high- magnification 45$^0$ views of MR surfaces fabricated with 400 radially polarized pulses of 1.12J/cm$^2$ per spot; (f) shows the top view of a single microstructure of the MR surface; (g,h) show low- and high- magnification 45$^0$ views of LR surfaces fabricated with 600 azimuthally polarized pulses of 0.42J/cm$^2$ per spot; (i) shows the top view of a single microstructure of the LR surface.

Using this approach, various complex surfaces can be produced upon different combinations of φ and NP, using radial and azimuthal CV beams respectively. Typical examples are presented in Fig. 6, however the variety of dual-scale biomimetic surfaces that can be produced is practically endless. Based on the sharpness of the principal microstructure, such surfaces can be classified to be of high (HR) (Fig. 6B a,b), medium (MR) (Fig. 6B c,d) and low (LR) (Fig. 6B e,f) roughness.

Natural and artificial multi-scale surfaces show extreme wetting characteristics that can be exploited for a vast range of applications, including microfluidics, friction reduction and lab-on-a-chip devices [2,5–7]. Direct laser micro/nanostructuring has been proven to be a unique tool to fabricate superhydrophobic and water-repellent biomimetic artificial surfaces [4]. In this context, the dual-rough biomimetic surfaces, described above, could provide an appropriate template for the observation of extreme wetting properties. To test this hypothesis the wetting properties of, 16mm$^2$, HR, MR and LR surfaces were evaluated. Following the fs-laser irradiation process, all surfaces showed superhydrophilic behavior (i.e., contact angles CA ~ 0°), which lasted for 6-7 days, depending on the surface structure, while the samples were stored in ambient conditions. The observed decrease of superhydrophilic behavior is attributed to the oxidation of the surface due to irradiation in air environment [33] and the removal of water molecules [34]. It is observed that the CA progressively increases to a terminal value after 9-10 days of storage in ambient air. This unusual behavior was observed before [33,34] but is not completely understood yet. It may be attributed to the adsorption of hydrocarbon molecules on the irradiated surface from the atmosphere [34] and/or a change of surface polarity [33]. The study of the water repellent and self-cleaning properties of the fabricated dual-scale surfaces is of special interest and measurements of the water repellent properties are currently under progress.

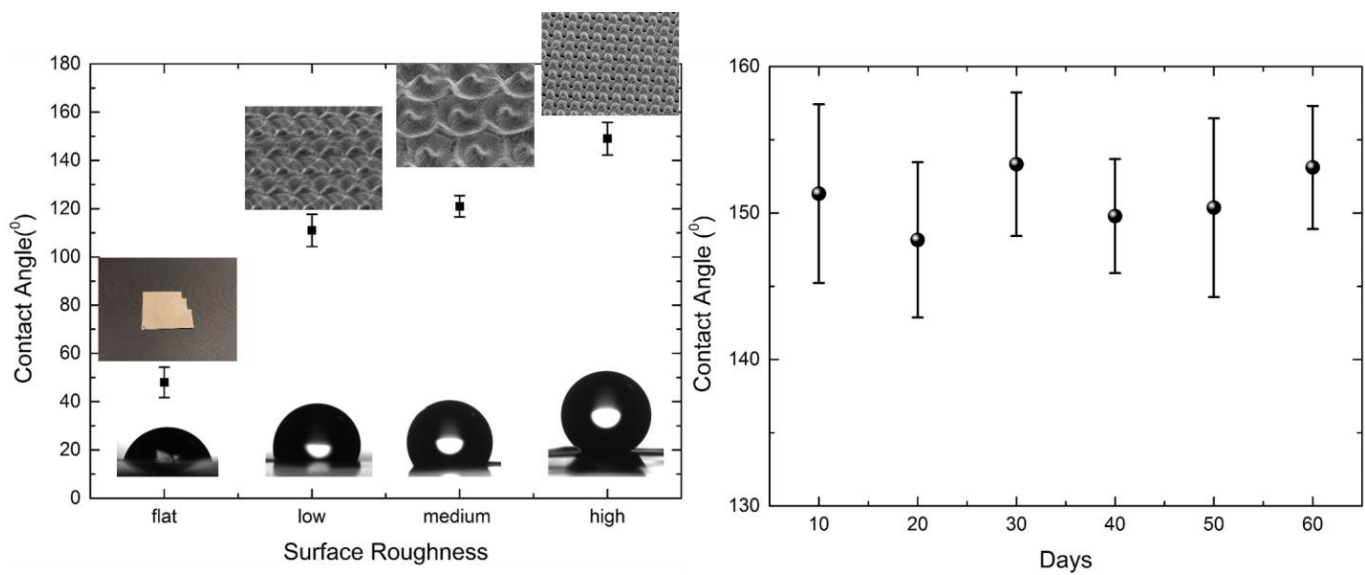

**Figure 7.** (a) Terminated CA values measured on the laser-fabricated low- (LR), medium- (MR) and high- (HR) roughness surfaces, exhibiting the morphology depicted in the insets. The CA of the non-irradiated, flat, Ni surface is also shown for comparison. The respective droplet profiles are also shown as insets; (b) Evolution of the terminal CA measured on the laser fabricated on the HR surface as a function of the time of storage in ambient conditions.

The terminated CA measured for water droplets placed on the three different types of dual-rough surfaces (LR, MR, and HR) is presented Fig. 7, along with a reference measurement for the non-irradiated flat surface. All measurements were performed 18-20 days after the fs-laser processing. In particular LR surfaces showed a CA of 111° ± 7°, MR structures exhibited CA of 122° ± 4°, while the HR surfaces had a CA value of 150° ± 5°. The CA of the non-irradiated (flat) surface, 48.6° ± 6°, is also shown for comparison. The CA measurements reveal a remarkable variation among the different surface morphologies, indicating that as the surface roughness increases the CA significantly increases as well. More importantly, the dual-rough HR morphology clearly exhibits a superhydrophobic nature, similar to the behavior exhibited by the natural suprhydrophobic archetype [1,2,4], i.e. the *Lotus* leaf surface. The terminal CA measured for the most hydrophobic HR surface was measured to be stable in time upon storage in ambient air for more than two months (Fig. 7b). It can be concluded that ultrafast laser structuring with CV beams can provide a unique tool for controlled structuring at different lengthscales, suitable to obtain biomimetic dual-scale surfaces with extreme wettability. The simplicity of the structuring process, together with the capability to fabricate a practically endless number of complex biomimetic structures, renders this technique very useful for tailoring the wetting response of surfaces in various wettability applications.

**Conclusions**

To summarize, we have presented a novel strategy to fabricate highly ordered, multi-directional and complex, biomimetic structures, via exploiting the unique and versatile angular profile of CV fs laser beams. Biomimetic surface structuring was realized through spot-by-spot and large area-scanning of radially and azimuthally polarized beams, giving rise to dual-scale, *Lotus* leaf-like, superhydrophobic surfaces as well as *Shark* skin-like biomimetic morphologies. Although the fabrication of these particular morphologies was demonstrated, laser processing with CV beams has a great potential to provide a plethora of complex structures and biomimetic surfaces. Our approach brings about a new concept in ultrafast laser structuring of materials and can be considered as an emerging laser based fabrication technique, which can be exploited for expanding the breadth and novelty of potential applications. No doubt, our approach requires further development before it can become a competitive technology. However, the wealth of arising possibilities in ultrafast laser based micro and nanofabrication prescribe a future where control of complex surfaces and subsequent functionality can be accomplished with a level of sophistication that we cannot presently envisage.

**Acknowledgments**

This work was supported by the European Horizon 2020 - FET OPEN Program 'LiNaBioFluid' under Grant No. 665337 (URL: www.laserbiofluid.eu). The authors acknowledge Prof. P. G. Kazansky for kindly offering the S-waveplate. The authors also acknowledge the technical support on software development of Mr. Andreas Lemonis.

**References**

1. Stratakis, E. I. & Zorba, V. *Biomimetic Artificial Nanostructured Surfaces. Nanomaterials for the Life Sciences* **7,** (2010).
2. Stratakis, E., Ranella, a. & Fotakis, C. Biomimetic micro/nanostructured functional surfaces for microfluidic and tissue engineering applications. *Biomicrofluidics* **5,** 1–31 (2011).
3. Yong, J., Chen, F., Yang, Q. & Hou, X. Femtosecond laser controlled wettability of solid surfaces. *Soft Matter* (2015). doi:10.1039/C5SM02153G
4. Zorba, V. *et al.* Biomimetic Artificial Surfaces Quantitatively Reproduce the Water Repellency of a Lotus Leaf. *Adv. Mater.* **20,** 4049–4054 (2008).
5. Paradisanos, I. *et al.* Gradient induced liquid motion on laser structured black Si surfaces Gradient induced liquid motion on laser structured black Si surfaces. **111603,** 0–5 (2015).
6. Vorobyev, A. Y. & Guo, C. Direct femtosecond laser surface nano/microstructuring and its applications. *Laser Photon. Rev.* **7,** 385–407 (2013).

7. Bonse, J. *et al.* Femtosecond laser-induced periodic surface structures on steel and titanium alloy for tribological applications. *Appl. Phys. A* **117,** 103–110 (2014).

8. Lu, Y., Hua, M. & Liu, Z. The Biomimetic Shark Skin Optimization Design Method for Improving Lubrication Effect of Engineering Surface. *J. Tribol.* **136,** 0317031–3170313 (2014).

9. Wang, Z., Li, Y.-B., Bai, F., Wang, C.-W. & Zhao, Q.-Z. Angle-dependent lubricated tribological properties of stainless steel by femtosecond laser surface texturing. *Opt. Laser Technol.* **81,** 60–66 (2016).

10. Simitzi, C. *et al.* Laser fabricated discontinuous anisotropic microconical substrates as a new model scaffold to control the directionality of neuronal network outgrowth. *Biomaterials* **67,** 115–128 (2015).

11. Jiang, H. B. *et al.* Bioinspired few-layer graphene prepared by chemical vapor deposition on femtosecond laser-structured Cu foil. *Laser Photonics Rev.* **450,** 441–450 (2016).

12. Gräf, S. & Müller, F. A. Polarisation-dependent generation of fs-laser induced periodic surface structures. *Appl. Surf. Sci.* **331,** 150–155 (2015).

13. Tsibidis, G. D., Barberoglou, M., Loukakos, P. A., Stratakis, E. & Fotakis, C. Dynamics of ripple formation on silicon surfaces by ultrashort laser pulses in subablation conditions. *Phys. Rev. B* **86,** 115316 (2012).

14. Gregorčič, P., Sedlaček, M., Podgornik, B. & Reif, J. Formation of laser-induced periodic surface structures (LIPSS) on tool steel by multiple picosecond laser pulses of different polarizations. *Appl. Surf. Sci.* **387,** 698–706 (2016).

15. Ruiz de la Cruz, A., Lahoz, R., Siegel, J., de la Fuente, G. F. & Solis, J. High speed inscription of uniform, large-area laser-induced periodic surface structures in Cr films using a high repetition rate fs laser. *Opt. Lett.* **39,** 2491 (2014).

16. Bonse, J., Höhm, S., Rosenfeld, A. & Krüger, J. Sub-100-nm laser-induced periodic surface structures upon irradiation of titanium by Ti:sapphire femtosecond laser pulses in air. *Appl. Phys. A* **110,** 547–551 (2013).

17. Tsibidis, G. D., Fotakis, C. & Stratakis, E. From ripples to spikes: A hydrodynamical mechanism to

interpret femtosecond laser-induced self-assembled structures. *Phys. Rev. B - Condens. Matter Mater. Phys.* **92,** 1–6 (2015).

18. Schmidt, V. & Belegratis, M. R. *Laser technology in biomimetics : basics and applications*.

19. Weber, R. *et al.* Effects of radial and tangential polarization in laser material processing. *Phys. Procedia* **12,** 21–30 (2011).

20. Ouyang, J. *et al.* Tailored optical vector fields for ultrashort-pulse laser induced complex surface plasmon structuring. *Opt. Express* **23,** 12562 (2015).

21. Hnatovsky, C., Shvedov, V., Krolikowski, W. & Rode, A. Revealing Local Field Structure of Focused Ultrashort Pulses. *Phys. Rev. Lett.* **106,** 123901 (2011).

22. Chen, R.-P. *et al.* Structured caustic vector vortex optical field: manipulating optical angular momentum flux and polarization rotation. *Sci. Rep.* **5,** 10628 (2015).

23. Jin, Y. *et al.* Dynamic modulation of spatially structured polarization fields for real-time control of ultrafast laser-material interactions. *Opt. Express* **21,** 25333–25343 (2013).

24. JJ Nivas, J. *et al.* Femtosecond laser surface structuring of silicon with Gaussian and optical vortex beams. *Appl. Surf. Sci.* (2016). doi:10.1016/j.apsusc.2016.10.162

25. Nivas, J. J. J. *et al.* Direct Femtosecond Laser Surface Structuring with Optical Vortex Beams Generated by a q-plate. *Nat. Publ. Gr.* 1–12 (2015). doi:10.1038/srep17929

26. Allegre, O. J., Perrie, W., Edwardson, S. P., Dearden, G. & Watkins, K. G. Laser microprocessing of steel with radially and azimuthally polarized femtosecond vortex pulses. *J. Opt.* **14,** 85601 (2012).

27. Anoop, K. K. *et al.* Femtosecond laser surface structuring of silicon using optical vortex beams generated by a q-plate. *Appl. Phys. Lett.* **104,** 241604 (2014).

28. Anoop, K. K. *et al.* Direct femtosecond laser ablation of copper with an optical vortex beam. *J. Appl. Phys.* **116,** 113102 (2014).

29. Beresna, M., Gecevicius, M., Kazansky, P. G. & Gertus, T. Radially polarized optical vortex converter created by femtosecond laser nanostructuring of glass. *Appl. Phys. Lett.* **98,** 2–4 (2011).

30. Tsibidis, G. D., Skoulas, E. & Stratakis, E. Ripple formation on nickel irradiated with radially

polarized femtosecond beams. *Opt. Lett.* **40,** 5172 (2015).

31. Oeffner, J. & Lauder, G. V. The hydrodynamic function of shark skin and two biomimetic applications. *J. Exp. Biol.* **215,** 785–795 (2012).

32. Wen, L., Weaver, J. C. & Lauder, G. V. Biomimetic shark skin: design, fabrication and hydrodynamic function. *J. Exp. Biol.* **217,** 1656–1666 (2014).

33. Kietzig, A. M., Hatzikiriakos, S. G. & Englezos, P. Patterned superhydrophobic metallic surfaces. *Langmuir* **25,** 4821–4827 (2009).

34. Bizi-Bandoki, P., Valette, S., Audouard, E. & Benayoun, S. Time dependency of the hydrophilicity and hydrophobicity of metallic alloys subjected to femtosecond laser irradiations. *Appl. Surf. Sci.* **273,** 399–407 (2013).

# Supplementary Material

## Two-dimensional fast Fourier transform on SEM images

In order to be able to extract spatial frequency information a 2D fast Fourier transform (2D-FFT) transform was employed. High-resolution (1280x1024) SEM pictures had been transformed in reverse space images via a 2D-FFT algorithm. The new dimensions of the generated Fourier images are inversely proportional to *x* and *y* dimensions of the original image. Fig.1 presents a typical SEM image of an irradiated laser spot using azimuthal polarization. While Fig.1S(c) shows the corresponding Fourier space image. The orange line represents the direction vertical to the ripple nanostructure. Along this direction the Fourier transformation detects a periodical fluctuation of the frequency intensity. This fluctuation exhibits an average frequency which is inversely proportional to the average ripple period.

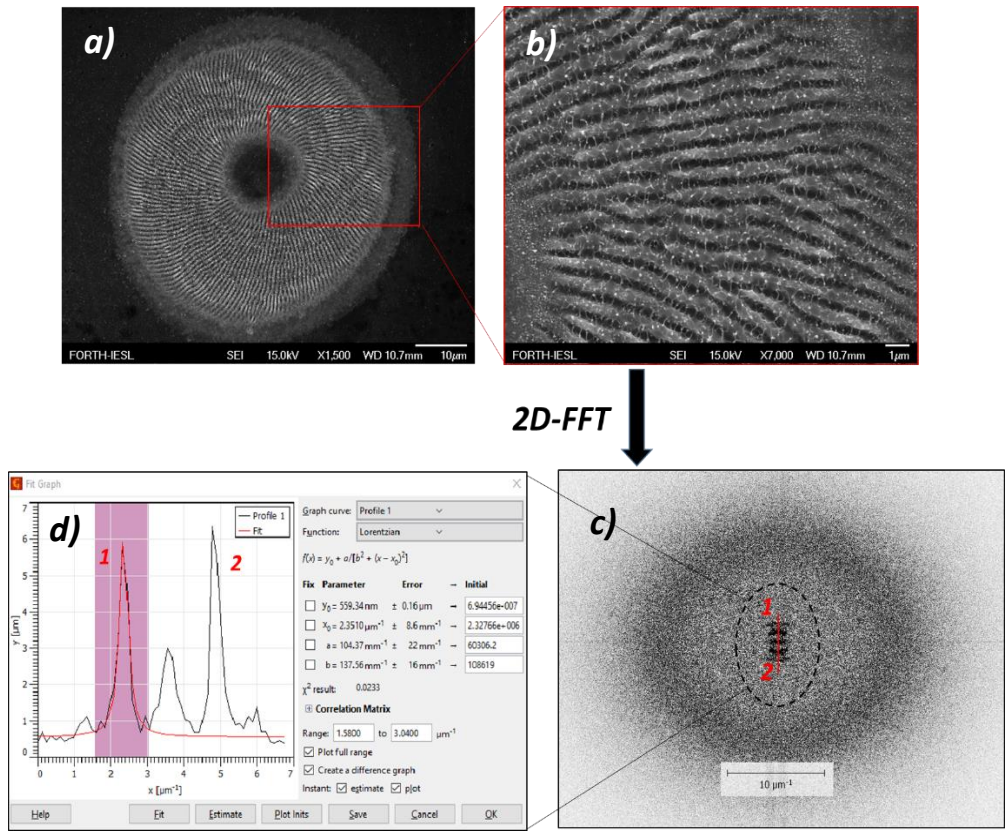

*Figure 1S: SEM images of ripple formation on nickel surface, (a) Laser spot, (b) higher magnification of the red box on (a). Image (c) represents the 2D-FFT of the image (b) without the labels on the bottom. The cross section of the black dashed ellipse of the Fourier space image (c) in presented on image (d) with peaks 1, 2 to correspond on the intensity fluctuation of the Fourier image.*

In particular, the distance between the centre of Fig.1S (d) and the first peak represents the characteristic frequency *f* of the periodic structure. In order to calculate the periodicity, $\Lambda$, of the structures first we calculate the average frequency of *1* and *2* peaks for a vertical as well as a horizontal image cross section (Fig. 1S(d)), and then the average period is given by the relation

$$<\Lambda> = 1/f.$$

Given that the beam can be changed from Gaussian to CV beam, which radically alters the spot surface profile, the LIPSS periodicity values and their relative errors using SEM images of three irradiated spots, produced with identical conditions were calculated. For the estimation of range of frequencies involved into the respective 2D-FFT images, we applied a Lorenzian fit on both peaks of the cross section and the error of each measurement is calculated using the following relation:

$$\Delta\Lambda = \left|-\frac{1}{f^2}\right|\Delta f \qquad (1)$$

were $\Delta f$ is the mean of the line widths for the two Lorenzian fit curves of the 2D-FFT image profile peaks.

## Initial Roughness

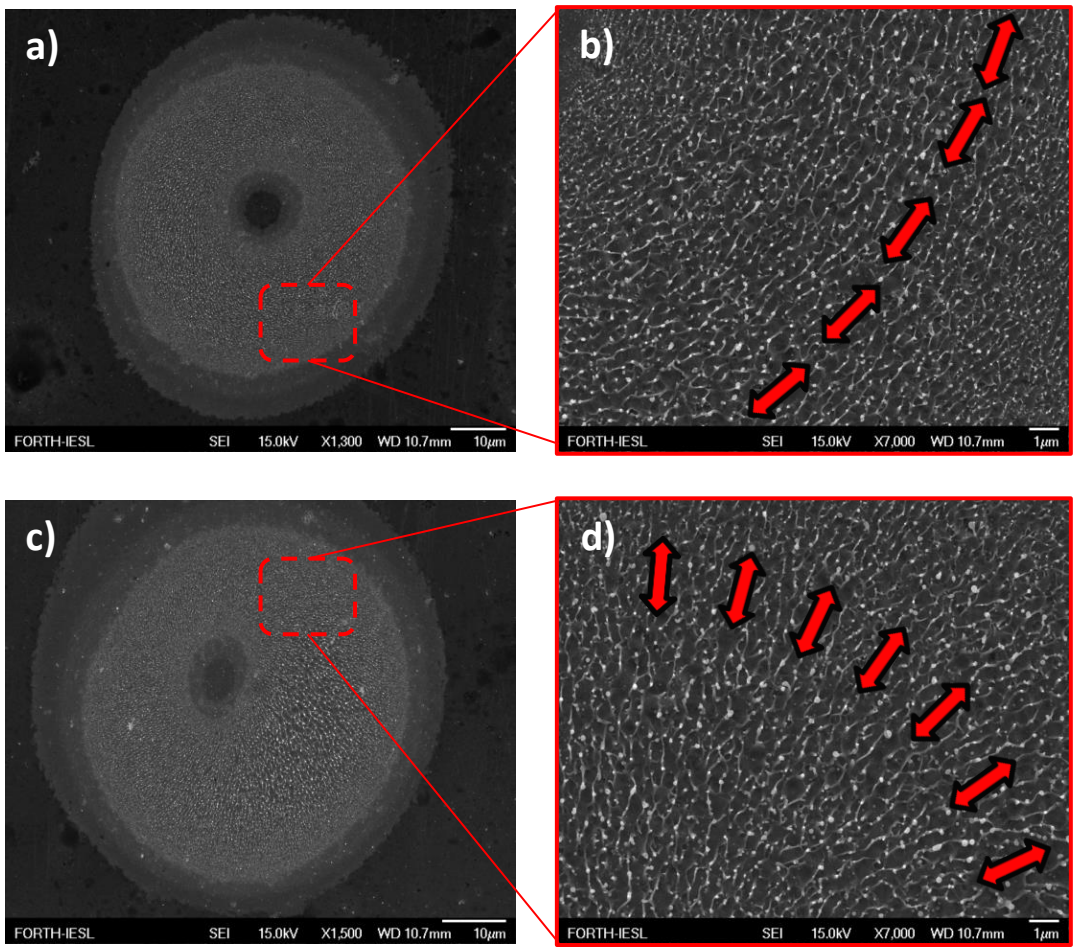

*Figure.2S: Figure shows SEM images of 0.49 J/cm$^2$ with azimuthal (a) 5 pulses, (b) higher magnification of the red dashed area. Radial polarization, (c) 5 pulses, (d) higher magnification of the red dashed area. The red arrows shows the electric field vector distribution.*

Results on SEM images showed that at low number of pulses (NP=2-10) and fluence values that range close to the ablation threshold, i.e. $0.17 J/cm^2$-$1.12 J/cm^2$, the surface shows a mushroom-like nano-roughness, with nanostructures aligned parallel to the incident electric field. Such nanostructures exhibit an average period of 100nm-250nm. Fig.2S shows SEM images of fs laser-iradiated spots at NP=5 and fluence of $φ=0.49 J/cm^2$ with azimuthal (Fig.2S(a),(b)) and radial polarization (Fig.2S(c),(d)) respectively. Ripple's formation is established following irradiation with NP=10 pulses.

Ripples where observed to always be perpendicular to the incident polarization, regardless the polarization state. Indeed, linear polarization produced ripple structures linearly aligned and perpendicular to the incident electrical field distribution. On the other hand azimuthal and radial polarization showed curved ripple structures, always perpendicularly arranged to the incident electric field. Consequently, ripples produced with azimuthal polarization showed radial orientation, while the ripple structures fabricated with radially distributed electrical field showed concentric circle-like symmetry.

# Inducing transparency on thin metallic films

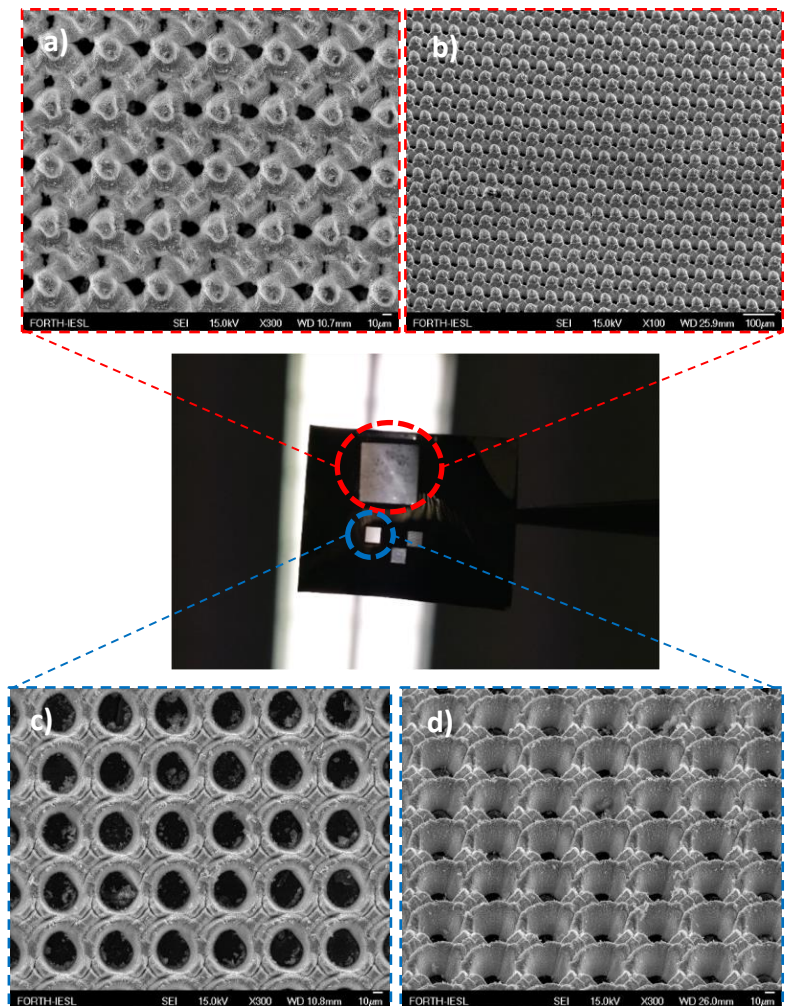

*Figure 3S: Photograph of Nickel films under light (middle). SEM images of semi-transparent areas fabricated with intense laser pulses. Images (a) top view (b) tilted at $45^0$ represents the red dashed highlighted area fabricated with CV beam and radial polarization at 500 pulses $9,45 J/cm^2$. Images (c) top view (d) tilted at $45^0$ represents the blue dashed highlighted area fabricated with Gaussian beam*

Irradiation of thin metallic films, with an average thickness of *d≃100μm*, using intense femtosecond pulses, at high fluence values, leads to massive material removal. Due to the small thickness of the film, the material removal could enhance the film transparency. In this context, we have conducted a series of experiments aiming at altering the layer thickness and fabricate laser structured transparent metallic membrane areas.

For the fabrication of the transparent membrane-like surfaces we used linearly polarized Gaussian as well as radially polarized CV beams. Typical SEM images are presented in Fig.3S. In the same Figure one can observe the treated areas of 4x4mm and 1x1mm under normal light illumination conditions. All laser treated areas show a significantly stronger light transmission compared to the untreated ones.

## Surface morphological profile

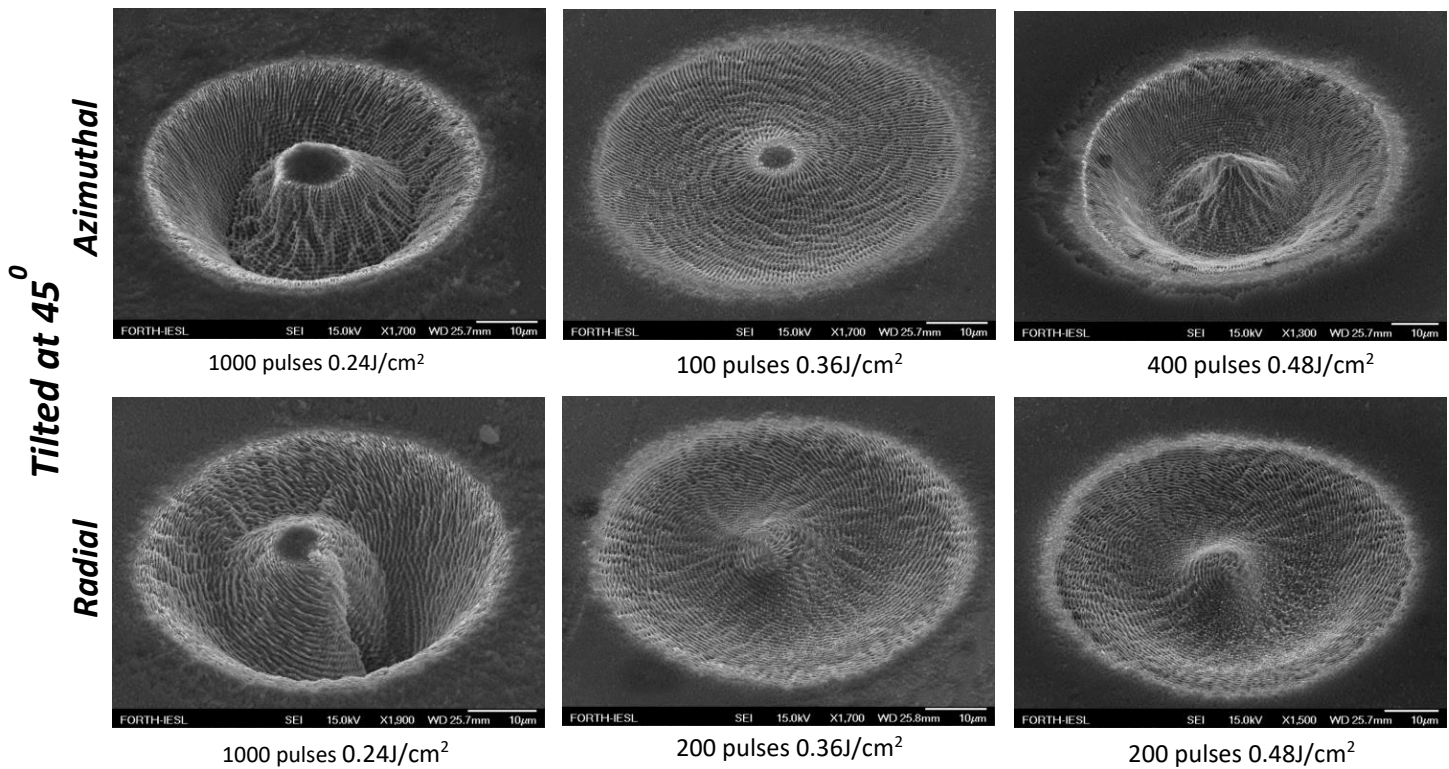

*Figure 4S: SEM images tilted at 45⁰ with the morphological profile on Ni surfaces after fs-irradiation with variable fluence and NP values.*

## CV & Gaussian beam profiles

Fig 5S. Presents the two laser beam images and their 3D plots. The laser beam profile was extracted with the use of a CMOS camera close to the focal plane for an S-linearly polarized Gaussian and a radially polarized CV beam. 3D plots were constructed with the pixel intensity values showing the spatial intensity variations for each case.

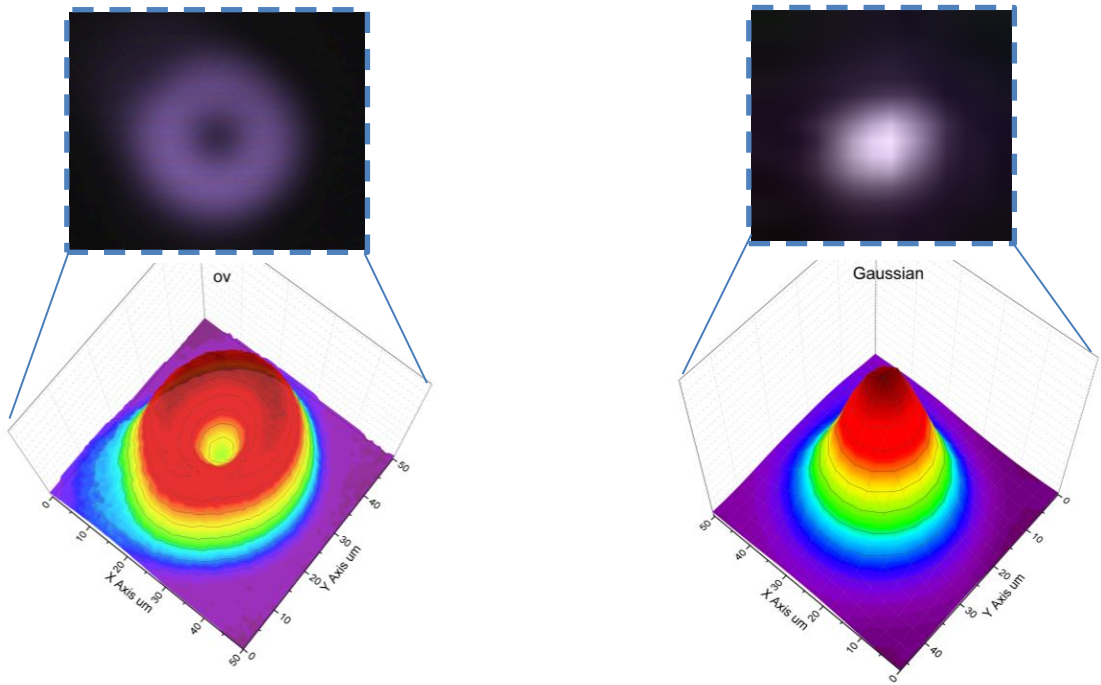

*Figure 5S: CV beam (left) and Gaussian beam (right).*

# Parametric with CV beam Line Scanning

## SEM Radial & Azimuthal lines at variable scan velocities

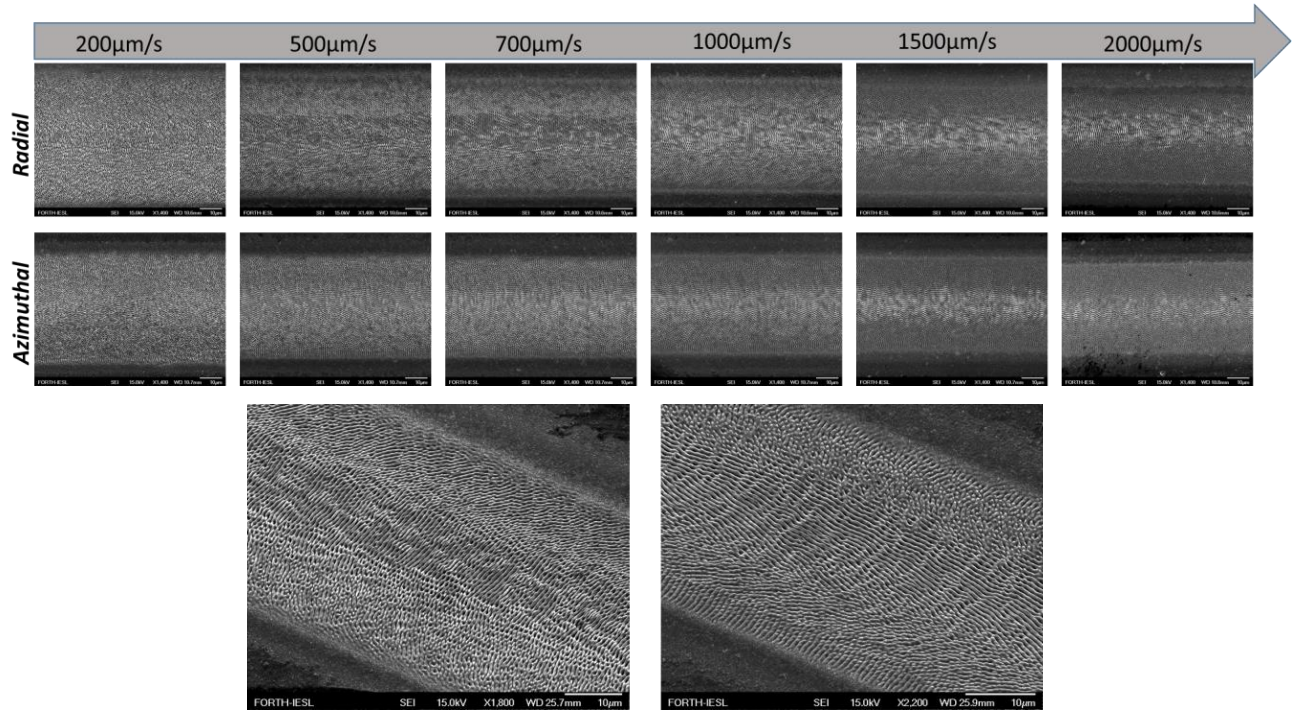

*Figure.6S: Line scans of φ=0.24cm² for 5 different scan velocity values.*

# Ablation threshold fluence estimation

Parametric study was initially conducted by performing single shot irradiations (NP=1), on nickel surfaces, at different fluences. The estimation of the ablation threshold fluence, was found to be at $\varphi_{th}=0.17 J/cm^2$ and $0.99 J/cm^2$ for the radially and azimuthally polarized CV beams respectively. It was aslo found that irradiation with fluence $0.11 J/cm^2 \leq \varphi \leq 0.17 J/cm^2$ can cause phase transition and give rise to a sort of surface roughness due to rapid resolidification of the melted material.

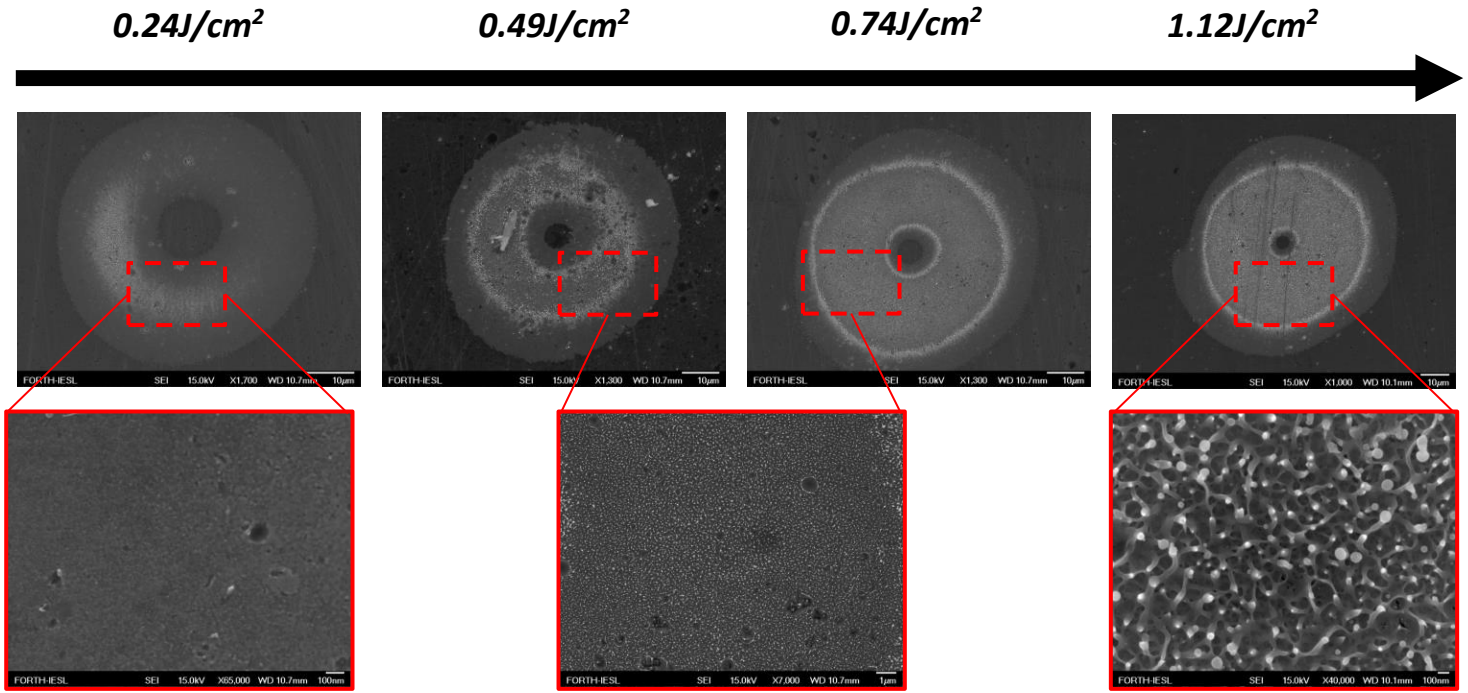

*Figure 1S: SEM images are presented from single shot (NP=1) irradiations of Nickel targets with azimuthally polarized laser pulses, for the estimation of the ablation threshold fluence area. The red outlined images represent higher magnification of the red dashed area.*

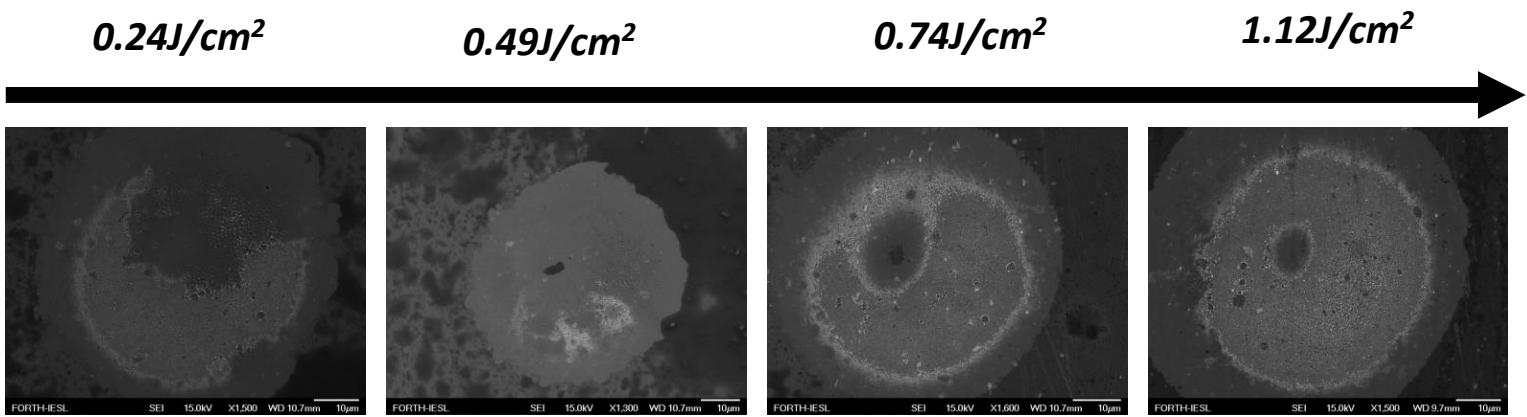

*Figure 2S: SEM images are presented from single shot (NP=1) irradiations of Nickel targets with radially polarized laser pulses, for the estimation of the ablation threshold fluence area.*

Consequently, we can identify two fluence value regimes, the sub-ablation one (0.11J/cm$^2$ ≤ φ ≤ 0.17J/cm$^2$) and the above-ablation one with fluences of 0.17J/cm$^2$ and higher. Fig.7 and Fig.8 present SEM images with single shot irradiation of Nickel surfaces with azimuthal and radial polarization beams respectively. It was observed that sub-ablation fluence values could not produce LIPSS at low pulse numbers, while above-ablation fluence values strongly ablate the surface for high number of pulses. In view of this we have decided to work at a fluence area that can give rise to LIPSS for both low and high number of pulses.